\documentclass[journal,twoside]{IEEEtran}

\usepackage{enumerate}
\usepackage{mathtools}
\usepackage{microtype}

\usepackage{pdflscape}

\usepackage{graphicx}
\usepackage{hyperref}
\usepackage{dblfloatfix}
\usepackage{rotating}

\usepackage{xcolor}
\usepackage{subcaption}
\usepackage[ruled,vlined]{algorithm2e}
\usepackage[T1]{fontenc}
\usepackage{lscape}
\usepackage[autostyle]{csquotes}
\usepackage{bbding}
\usepackage{todonotes}
\usepackage{smartdiagram}
\usepackage{comment}
\usepackage{adjustbox}
\usepackage{tikz}
\usepackage{amsmath}
\usepackage{multirow}

\usepackage{amssymb}
\usepackage{pifont}

\DeclareUnicodeCharacter{2212}{-}

\usepackage{comment}
\def\BibTeX{{\rm B\kern-.05em{\sc i\kern-.025em b}\kern-.08em
    T\kern-.1667em\lower.7ex\hbox{E}\kern-.125emX}}

\ifCLASSINFOpdf
\else
 
\fi

\usepackage[a4paper, total={180mm,257mm}]{geometry}

\begin{document}

%


\title{Novel Area-Efficient and Flexible Architectures for Optimal Ate Pairing on FPGA}


\author{\IEEEauthorblockN{Oussama Azzouzi\IEEEauthorrefmark{1},
Mohamed Anane\IEEEauthorrefmark{2}, 
Mouloud Koudil\IEEEauthorrefmark{2}, 
Mohamed Issad\IEEEauthorrefmark{3},
Yassine Himeur\IEEEauthorrefmark{4},
}\\
\IEEEauthorblockA{\IEEEauthorrefmark{1}
Ecole Nationale Supérieure d’Informatique, Laboratoire des Méthodes de Conception des Système, BP 68M, 16309, Oued-Smar, Alger, Algérie (o\_azzouzi@esi.dz)}\\
\IEEEauthorblockA{\IEEEauthorrefmark{2}Centre Universitaire El Cherif Bouchoucha d’Aflou, Laghouat, Algérie}\\
\IEEEauthorblockA{\IEEEauthorrefmark{3}Department of System and Multimedia Architecture, Centre de Développement des Technologies Avancées (CDTA), Algiers, Algeria}\\
\IEEEauthorblockA{\IEEEauthorrefmark{4}College of Engineering and Information Technology, University of Dubai, Dubai, UAE (yhimeur@ud.ac.ae)}\\
}


\maketitle

\begin{abstract}
While FPGA is a suitable platform for implementing cryptographic algorithms, there are several challenges associated with implementing Optimal Ate pairing on FPGA, such as security, limited computing resources, and high power consumption. To overcome these issues, this study introduces three approaches that can execute the optimal Ate pairing on Barreto-Naehrig curves using Jacobean coordinates with the goal of reaching 128-bit security on the Genesys board. The first approach is a pure software implementation utilizing the MicroBlaze processor. The second involves a combination of software and hardware, with key operations in $F_{p}$ and $F_{p^{2}}$ being transformed into IP cores for the MicroBlaze. The third approach builds on the second by incorporating parallelism to improve the pairing process. The utilization of multiple MicroBlaze processors within a single system offers both versatility and parallelism to speed up pairing calculations. A variety of methods and parameters are used to optimize the pairing computation, including Montgomery modular multiplication, the Karatsuba method, Jacobean coordinates, the Complex squaring method, sparse multiplication, squaring in $G_{\phi 6}F_{p^{12}}$, and the addition chain method. The proposed systems are designed to efficiently utilize limited resources in restricted environments, while still completing tasks in a timely manner.
\end{abstract}

\begin{IEEEkeywords}
Optimal Ate pairing,d Flexible architecture, Virtex-5, MicroBlaze, Montgomery modular multiplication, Karatsuba method.
\end{IEEEkeywords}

\IEEEpeerreviewmaketitle
\section{Introduction}
Cryptography is a crucial technology for ensuring the security and privacy of data in today's digital world \cite{rathore2022novel,himeur2018robust}. Cryptography is the practice of converting plain text into a coded message to protect it from unauthorized access or tampering. It plays a crucial role in securing communication channels and protecting sensitive information, such as financial transactions, personal information, and state secrets \cite{can2023comprehensive,himeur2016performance}. Cryptography is widely used in various applications, including online banking, e-commerce, secure communication between individuals and organizations, and in the protection of critical infrastructure systems \cite{cong2022fpga,himeur2022latest}. Without cryptography, sensitive information would be vulnerable to cyber-attacks and malicious activities, leading to severe consequences such as data breaches, identity theft, and financial loss \cite{ullah2023elliptic,sayed2021intelligent}. Therefore, cryptography is essential for ensuring the confidentiality, integrity, and availability of data and maintaining trust in the digital world \cite{dizon2023value,ullah2023elliptic}. On the other hand, field-programmable gate arrays (FPGAs) are increasingly being utilized in edge computing environments due to their versatility, configurability, and performance advantages \cite{faj2023scalable,alsalemi2021smart}. In edge computing, FPGAs play a crucial role in accelerating data processing, enabling real-time analytics, and enhancing overall system performance \cite{haghi2023flash}.

The concept of pairing functions was introduced by André Weil in 1948, and later it was utilized in cryptography with the employment of elliptic curve bilinear pairings. The bilinear property enables the transformation of the discrete logarithm issue from an elliptic curve to the finite field $F_p$. This brought about the emergence of the MOV attack \cite{menezes1991reducing} and Frey-Rück attack \cite{frey1994remark}. The widespread use of pairing functions in cryptography emerged in the early 2000s after Joux introduced the tripartite key exchange scheme for Diffie-Hellman \cite{joux2000one}. Since then, pairing functions have been implemented in a variety of advanced cryptosystems including identity-based signatures \cite{zhou2022efficient}, searchable encryption \cite{andola2022searchable}, and functional encryption \cite{mera2022efficient}. One of the most prominent applications of pairing functions is the Identity-Based Encryption (IBE) \cite{boneh2001identity} proposed by Boneh and Franklin.

Pairing functions, which are used in cryptography, are typically constructed using a combination of the Miller Loop and Final Exponentiation. The performance of these functions is dependent on the arithmetic used in the primary field $F_p$ and its extensions $F_{p^{k}}$. To improve the efficiency of pairings, various curves have been discovered that offer improved computation and enhanced security. Freeman and colleagues provide a comprehensive categorization of such "pairing-friendly" curves in their work \cite{freeman2010taxonomy}. Currently, one of the most favorable options for computational efficiency and security is the use of Barreto-Naehrig (BN) curves \cite{barreto2005pairing}. Many articles have been published that propose protocols utilizing pairings \cite{joux2000one,boneh2001identity}, while others focus on improving the computation of pairings \cite{vercauteren2009optimal,scott2009final}. A smaller number of articles propose FPGA implementations for computing pairing functions \cite{bahadori2020compact,oussama2019software,ghosh2012secure,hao2016dual}. Recently, there has been a growing interest in implementing cryptographic pairings, with hardware implementations being considered a superior approach compared to software developments. Examples of advancements in pairing function implementations in cryptography include the following studies: In 2010, Ghosh et al. were the first to implement pairing functions on BN-curves, offering 128-bit security \cite{ghosh2010high}. 
Moving on, Cheung et al. \cite{cheung2011fpga} improved execution time in their solution for optimal Ate pairing with 126-bit security by adopting the Residue Number System. Fan et al. provided a hardware implementation for pairing \cite{fan2011efficient} in 2012 utilizing $F_p$-arithmetic. Then, using $F_{p^k}$-arithmetic in hardware, Ghosh et al. created a complete hardware implementation of Ate and optimal Ate pairing \cite{ghosh2012secure}. 
A method for computing the challenging portion of the final exponentiation with the least amount of resource consumption was introduced in 2015 by Duquesne et al \cite{duquesne2016memory}. A high-speed and effective optimal Ate pairing processor implementation over BN and BLS12 curves on FPGA was lastly proposed by Sghaier et al. in 2018 \cite{sghaier2018high}.

\color{black}
On the other hand, the emergence of post-quantum cryptography (PQC) and the utilization of alternative schemes like Kyber and Dilithium as replacements for RSA/ECC have generated significant interest due to their ability to withstand quantum attacks \cite{bernstein2017post}. An ideal choice for low-resource applications is the ECC since it offers the same level of security with smaller key sizes compared to other existing public key encryption schemes. 
An effective platform for an embedded co-processor is achieved by designing efficient functional units for elliptic curve computations over binary fields, making it suitable for low-resource applications.
\cite{koziel2015low} presents an efficient co-processor for elliptic curve cryptography (ECC) over binary Edwards curves, designed for area-constrained devices. By utilizing state-of-the-art binary Edwards curve equations, it achieves a secure yet fast implementation of point multiplication. The co-processor offers the same level of security as other public key encryption schemes but with smaller key sizes, making it ideal for low-resource applications. Synthesis results show that it requires about 50\% fewer clock cycles for point multiplication and occupies a similar silicon area compared to recent literature.

Although ECC is widely implemented and efficient, its security is reliant on the complexity of the elliptic curve discrete logarithm problem, which can be solved by quantum computers employing Shor's algorithm. To ensure long-term security, researchers have actively explored and developed PQC schemes that offer robust protection against the threats posed by quantum computing \cite{ullah2023elliptic}. Kyber, focusing on key exchange protocols, and Dilithium, specializing in digital signatures, exemplify such schemes. The adoption of post-quantum algorithms such as Kyber and Dilithium represents a proactive approach in guaranteeing the ongoing security of cryptographic systems in anticipation of forthcoming advancements in quantum computing \cite{imran2023high}. 
%
For instance, 
the authors in \cite{jalali2017supersingular} demonstrate the practicality and efficiency of the Supersingular Isogeny Diffie-Hellman (SIDH) key exchange on 64-bit ARM architectures. SIDH is a cryptographic key exchange protocol that relies on supersingular isogenies, a concept derived from elliptic curve theory. Moving on, Anastasova et al. \cite{anastasova2021fast}  explore the fast strategies for the implementation of Supersingular Isogeny Key Encapsulation (SIKE) Round 3 on ARM Cortex-M4, showcasing optimized techniques. Additionally, \cite{sarker2020error} discusses error detection architectures for Ring Polynomial Multiplication and Modular Reduction of Ring-LWE, providing valuable insights into ASIC implementations. These works collectively contribute to the advancement of cryptographic implementations on resource-constrained platforms and are crucial in the context of secure and reliable systems. Besides, \cite{bisheh2021cryptographic} investigates hardware accelerators that are specifically designed to improve the efficiency of digital signature operations utilizing the Ed25519 algorithm. Ed25519 is a widely employed digital signature algorithm that relies on the elliptic curve Curve25519.
\color{black}

\color{black}
Moving on, to acknowledge the significance of lightweight cryptography (LWC) and building blocks in low-energy and low-power implementations, many studies have been proposed in the literature. For instance, \cite{bayat2013dual} presents low-complexity superserial architectures for dual basis (DB) multiplication over GF(2m) to achieve lightweight cryptographic algorithms. It is the first time such a multiplier is proposed in open literature. Moving forward, \cite{subramanian2017reliable} explores cryptographic architectures' reliability in providing security properties to sensitive usage models. It considers two underlying block ciphers suitable for authenticated encryption algorithms: the Advanced Encryption Standard type and Feistel network structure. 
In the same direction, \cite{kermani2018reliable} discusses augmenting block ciphers' confidentiality with authentication using the standardized Galois Counter Mode (GCM). Existing GCM error detection methods are either limited to specific architectures or ineffective against biased faults. 
\color{black}

While FPGA is a suitable platform for implementing cryptographic algorithms, there are several challenges associated with implementing optimal Ate pairing on FPGA. Some of these challenges include (i) high computational complexity due to the fact that optimal Ate pairing involves complex mathematical operations; (ii) \textcolor{black}{lightweight cryptography \cite{bayat2013dual} poses resource constraints on FPGAs, necessitating the optimization of optimal Ate pairing to efficiently utilize logic gates, memory, and power. Techniques like algorithmic optimization, parallelization, and hardware-specific optimization can enable faster and more efficient FPGA implementations \cite{canto2023algorithmic}}; (iii) high power consumption as optimal Ate pairing requires a large number of clock cycles to execute, which increases the power consumption of the FPGA \cite{liu2022fpga}; and (iv) design complexity which is due to the requirement of a thorough understanding of the mathematical operations involved, as well as the hardware design and implementation \cite{liu2022design}.
\textcolor{black}{(v) FPGA implementations are vulnerable to physical attacks due to the hardware's inherent properties. Techniques such as resistance to power analysis, and secure key storage must be employed to mitigate vulnerabilities related to side-channel attacks \cite{kaur2023comprehensive}. The assessment of combined attacks requires a deep understanding of potential vulnerabilities in FPGA designs, the detection mechanisms employed by attackers, and techniques for analyzing power consumption. Implementation of specific countermeasures is possible, including the utilization of error detection and correction techniques to identify and mitigate injected faults. Furthermore, reducing information leakage through masking techniques and continuously monitoring power consumption to detect anomalies can be employed.}

In this paper, we propose three different methods for implementing optimal Ate pairing on BN-curves with 128-bit security using the Virtex-5 circuit. Our first method is a full software implementation on an FPGA with a MicroBlaze processor, offering high flexibility. Our second approach integrates an intellectual property (IP) core written in VHDL into the MicroBlaze, offering a balance of flexibility, area, and speed. The third method builds upon the second by utilizing parallelism for enhanced computation speed. Our work adds to the existing literature on FPGA-based pairing implementations by providing flexible solutions that support various pairing methods and parameters, such as Montgomery modular multiplication, the Karatsuba method, and the addition chain method. The goal is to minimize resource consumption while maintaining reasonable execution times by combining a mixed software and hardware approach and utilizing parallelism. Overall the main contributios of this paper are summarized as follows:

\begin{itemize}
\item Proposing three methods for implementing optimal Ate pairing on BN-curves with 128-bit security using Virtex-5 circuit by (i) using full software implementation on FPGA with MicroBlaze processor, offering high flexibility; (ii) introducing IP core written in VHDL integrated into MicroBlaze, offering balance of flexibility, area, and speed; and (iii) building on second method by utilizing parallelism for enhanced computation speed.
\item Adding to existing literature on FPGA-based pairing implementations.
\item Providing flexible solutions that support various pairing methods and parameters (Montgomery modular multiplication, Karatsuba method, addition chain method)
\item Helping minimizing resource consumption while maintaining reasonable execution times through a mixed software and hardware approach and utilization of parallelism
\end{itemize}

The reminder of this paper is organized as follows. Section 2 presents an overview of optimal Ate pairing on BN curves and the relevant parameters. Section 3 covers the IP cores made using VHDL. Section 4 details three methods for embedding optimal Ate pairing on FPGA. In Section 5, our implementation results are evaluated and compared to previous studies. Finally, Section 6 concludes the research findings.

\section{Optimal Ate Pairing over BN-Curves}

A pairing function, denoted as $e(P,Q)$, maps two points, $P$ and $Q$, on an elliptic curve $E$ to an element in an extension field $F_{p^{12}}$ for two cyclic additive groups $G_1$ and $G_2$ and a multiplicative group $G_3$. It is required to possess the properties of bilinearity and non-degeneracy. One of the most useful properties derived from bilinearity is : $for \; P\in G_1$, $Q\in G_2$, we have:

\begin{equation}
\forall  j\in \mathbb{N} : e([j]P,Q) = {e(P,Q)}^j = e(P,[j]Q) 
\end{equation}

As stated in \cite{barreto2005pairing}, Barreto-Naehrig introduced a technique for creating pairing-friendly elliptic curves that are defined over a prime field $F_p$. These curves, known as ordinary elliptic curves, are crucial for achieving a 128-bit level of security and for efficient pairing computation. They are defined by the following equation:

\begin{equation}
E:  {y^2 = x^3 + b} \;\; where \; b\neq 0
\end{equation}

The embedding degree for BN-curves is 12. Additionally, the prime field characteristic, $p$, the group order, $r$, and the trace of Frobenius, $t_r$ of these curves are determined by the following:

\begin{equation}
\begin{array}{c}
p(t) =36t^{4}+36t^{3}+24t^{2}+6t+1 \\ 
r(t) =36t^{4}+36t^{3}+18t^{2}+6t+1 \\ 
t_{r}(t) =6t^{2}+1,\;where\;t\in \mathbb{Z}~~~~~~~~~~
\end{array}%
\end{equation}

The choice of parameters plays a crucial role in the security and efficiency of the pairing function. The variable $t$ is chosen so that both $p$ and $r$ are prime numbers. Furthermore, it is important to select a large enough value of $t$ in order to attain a higher level of security. According to the recommendations of National Institute of Standards and Technology (NIST) \cite{barker2007nist}, for a security level similar to AES 128 bits, $t$ should be such that $log_2(r(t)) \geq 256$ and $3000\leq k.log_2(p(t)) \leq 5000$ , which leads to $t$ having roughly 64 bits.

The notation $E[r]$ represents the $r$-torsion subgroup of $E$, and $\pi_p$ is the Frobenius endomorphism that maps $E$ to $E$, defined as $\pi_p(x,y)=(x^p,y^p)$. We define $G_1$ as $E(F_p)$, $G_2$ as a subset of $E(F_{p^{12}})$, and $G_3$ as $\mu_r$ which is part of $F_{p^{12}}^\ast$. The optimal Ate pairing on BN-curves can be represented by the following mapping:

\begin{equation}
\begin{array}{c}
e_{opt}:G_{2}\times G_{1}\rightarrow G_{3} \\ 
(Q,P)\mapsto (f_{s,Q}(P)\;.\;f_{[s]Q,\pi _{p}(Q)}(P)\;.\;f_{[s]Q+\pi
_{p}(Q),} \\ 
-\pi _{p}^{2}(Q)(P))^{\frac{p^{12}-1}{r}}%
\end{array}%
\end{equation}

The optimal Ate pairing algorithm, as described in \cite{vercauteren2009optimal}, is outlined in Algorithm \ref{alg01}. Using the non-adjacent form (NAF representation), the algorithm has three main steps. The Miller Loop, computed in lines 3-11, generates the value of $f_{s,Q}(P)$. Point additions with the Frobenius map of point $Q$ are calculated in lines 12-14, and the final exponentiation is performed in line 15. Note that in this algorithm, $s$ is defined as $6t+2$.

\begin{algorithm}
\caption{Optimal Ate pairing over BN-curves}
\label{alg01}
\KwData{$P \in G_1 \;$ and $\; Q \in G_2$}
\KwResult{$a_opt(Q,P)$}
\SetAlgoLined

$write \;s=6t+2 \;$ as $s=\sum_{i=0}^{L-1} s_i 2^i \;$ ,where $s_i\in\{-1,0,1\}; \; L=bitlength(s) $ \\
$ T \gets Q; \;\; f\gets 1;$ \\

\For{$i \gets L-2$ \textbf{to} $0$} {
    $f \gets f^2.l_{T,T}(P); \;\; T \gets 2T;$ \\
    \If{$s_i=-1$}{
        $ f \gets f.l_{T,-Q}(P); \;\; T \gets T-Q;$\\
    }
    \If{$s_i=1$}{
        $ f \gets f.l_{T,Q}(P); \;\; T \gets T+Q;$\\
    }
}
$Q_1 \gets \pi_p(Q); \;\; Q_2 \gets \pi_{p^2}(Q);$ \\ 
$f \gets f.l_{T,Q_1}(P); \;\; T \gets T+Q_1;$ \\
$f \gets f.l_{T,-Q_2}(P); \;\; T \gets T-Q_2;$ \\ 
$f \gets f^{\frac{p^{12}-1}{r}};$  \\
$return \;f$;
\end{algorithm}

The key operations utilized in the optimal Ate pairing algorithm, as detailed in \cite{vercauteren2009optimal}, include: Doubling and Addition steps (occurring on lines 4, 6, 9, 13 and 14), Sparse multiplication as outlined in \cite{beuchat2010high} (on lines 4, 6, 9, 13 and 14) which is a multiplication in $F_{p^{12}}$ where the second operand has half of the coefficients equal to zero, the Frobenius operation (on line 12), Squaring in the cyclotomic subgroup $G_{\phi 6}(F_{p^{12}})$ (on line 15), and the Final Exponentiation (on line 15). The doubling and addition steps are executed in $F_{p^2}$, while most of the other operations are performed in $F_{p^{12}}$.

In order to efficiently perform extended field operations in $F_{p^{12}}$, advanced techniques can be used to construct the arithmetic step by step in smaller extensions fields, such as the polynomial irreducible $X^k-\beta$, and a tower of extensions of degree 2 and 3 can be utilized, similar to the method presented in \cite{joye2009software}.

\begin{equation}
\begin{array}{l}
F_{p^2} = \frac{F_p[\mu]}{\mu^2 - \beta}, \; where \; \beta=-5 \\
F_{p^6} = \frac{F_{p^2}[\nu]}{\nu^3 - \xi}, \; where \; \xi=\mu \\
F_{p^{12}} = \frac{F_{p^6}[\omega]}{(\omega^2 - \nu)} 
\end{array}%
\end{equation}

A technique for representing elements of the field $F_{p^{12}}$ using a combination of smaller extensions can be used to improve the speed of pairing. This method, called a "towering scheme," expresses an element $f$ as $f=g+h\omega$, where $g,h \in F_{p^6}$. The element $g$ can be further broken down into $g_0 + g_1 \nu + g_2 \nu^2$, and the same is done for $h$, where $g_i,h_i \in F_{p^2}$ for $i=0,1,2$. This approach, as outlined in \cite{joye2009software}, allows for a faster computation of pairing.

\subsection{Miller Loop}
The Miller algorithm, as found in popular pairings such as Weil, Tate, Ate and optimal Ate pairing \cite{miller2004weil}, is used to construct a rational function $f_{r,P}$ associated with a point $P$ on an elliptic curve $E$, which is evaluated at another point $Q$. This is achieved through an iterative process using the double and addition method, and the function $f_{r,P}$ is defined by its divisor.

\begin{equation}
Div(f_{r,P})=r(P) - ([r]P) - (r-1)(P_\infty)  
\end{equation}

where $r$ is an integer and $P_\infty$ denotes the point at infinity. The function is calculated by utilizing Miller's equality.

\begin{equation}
f_{[i+j],P}=f_{[i]P} . f_{[j]P} . \frac{l_{[i]P,[j]P}}{v_{[i+j]P}}
\end{equation}

where $l_{[i]P,[j]P}$ is the line passing through $[i]P$ and $[j]P$, and $v_{[i+j]P}$ is the vertical to $E$ at $[i+j]P$. The performance of the Miller Loop is affected by the number of bits in the exponent, as well as its Hamming weight. 

\subsubsection{Doubling and tangent equations}
The formulas for $T=2Q=(X_T,Y_T,Z_T)$ in Jacobian coordinates are defined as follows:

\begin{equation}
\begin{array}{l}
    X_R = 9X_T^4-8Y_T Y_T^2 \\
    Y_R = 3X_T^2(4X_T Y_T-X_R)-8Y_T^4 \\
    Z_R = 4X_T Y_T  
\end{array}%
\end{equation}

To find the tangent line equation at $T$ when a point $P=(x_p,y_p)$ in $E(F_p)$ is given in affine coordinates, the following calculation can be performed:

\begin{equation}
l_{T,T}(P)=(4Z_R Z_T^2 y_P)-(6X_T^2 Z_T^2 x_p)\omega+(6X_T^3-4Y_T^2)\omega^2 \in F_{p^{12}}
\end{equation}

\subsubsection{Addition and line equations}
The formulas for addition $R=T+Q=(X_R,Y_R,Z_R)$ are defined as follows:

\begin{equation}
\begin{array}{cc}
X_{R}= & (2Y_{Q}Z_{T}^{3}-2Y_{T})^{2}-4(X_{Q}Z_{T}^{2}-X_{T})^{3} \\ 
& -8(X_{Q}Z_{T}^{2}X_{T})^{2}X_{T} \\ 
Y_{R}= & (2Y_{Q}Z_{T}^{3}-2Y_{T})\;(4(X_{Q}Z_{T}^{2}-X_{T})^{2}X_{T}-X_{R})
\\ 
& -8Y_{T}(X_{Q}Z_{T}^{2}-X_{T}) \\ 
Z_{R}= & 2Z_{T}(X_{Q}Z_{T}^{2}-X_{T})%
\end{array}%
\end{equation}

The equation of the line passing through $T$ and $Q$ when evaluated at point $P$ is:

\begin{equation}
\begin{array}{c}
l_{T,Q}(P)=(4Z_{T}(X_{Q}Z_{T}^{2}-X_{T})y_{p})-(4x_{p}(Y_{Q}Z_{T}^{3}+Y_{T}))\omega +
\\ 
(4X_{Q}(Y_{Q}Z_{T}^{2}X_{Q}-Y_{T})-4Y_{Q}Z_{T}(X_{Q}Z_{T}^{2}-X_{T}))\omega
^{2}\in F_{p^{12}}%
\end{array}%
\end{equation}

After the Miller Loop has been completed, an additional step known as the Final Exponentiation must be performed. This step involves raising the result of the Miller Loop to the power $\frac{p^k-1}{r}$.

\subsection{Final Exponentiation}
Several techniques can be employed to perform the Final Exponentiation step in algorithm \ref{alg01}. The traditional approach is to use the square and multiply method, however, this method can be time-consuming as the exponent $e= \frac{p^{12}-1}{r}$ is large. To reduce computation time, the exponent can be broken down into smaller components.

\begin{equation}
e= \frac{p^{12}-1}{r} = (p^6-1).(p^2+1).\frac{p^4-p^2+1}{r}
\end{equation}

To calculate the first part $f^{(p^6-1)(p^2+1)} \in F_{p^{12}}$, which is the easy part, we can use simple conjugation and Frobenius operations to raise $f$ to the power $p^6$ and $p^2$, respectively. This results in an element of the cyclotomic subgroup $G_{\phi6} (F_{p^2})$. There are various methods available in the literature for calculating the hard part of the Final Exponentiation. One such method is the approach proposed by Scott et al. in 2009 \cite{scott2009final}, which is based on addition chain. This method simplifies computations by keeping all elements involved within the cyclotomic subgroup $G_{\phi6} (F_{p^2})$, reducing the number of required operations for $f^2$ computations \cite{granger2010faster}, and allowing for inversions to be performed as a simple conjugation \cite{beuchat2010high}.

The addition chain method utilizes the polynomial representation of $p$ and $r$ in $t$ to effectively decompose the hard part of the Final Exponentiation. This method involves a clever procedure that involves the computation of ten intermediate values, as follows:

\begin{equation}
f^t, f^{t^2}, f^{t^3}, f^p, f^{p^2}, f^{p^3}, f^{(tp)}, f^{(t^2 p)}, f^{(t^3 p)}, f^{(t^2 p^2)}
\end{equation}

These crucial components are employed to build a chain of multiplications, the evaluation of which results in the Final Exponentiation $f^e$, through the implementation of the following equation:

\begin{equation}
\begin{array}{c}
\lbrack f^{p}.f^{p^{2}}.f^{p^{3}}].{[\frac{1}{f}]}^{2}.{[(f^{t^{2}})^{p^{2}}]%
}^{6}.{[\frac{1}{(f^{t})^{p}}]}^{6}.{[\frac{1}{{(f^{t}.f^{t^{2}})}^{p}}]}%
^{18} \\ 
.{[\frac{1}{f^{t^{2}}}]}^{30}.{[\frac{1}{(f^{t^{2}}.f^{t^{3}})^{p}}]}^{36}%
\end{array}%
\end{equation}

To raise an element to the power $p$, we can compute it by applying the Frobenius operation. Additionally, to raise an element to the power $t$, which can be time-consuming, we can use the square and multiply method. Lastly, we can use Fermat's little theorem to perform modular inversion in $F_p$ by using this equation:

\begin{equation}
A^{-1} \equiv A^{p-2} mod p 
\end{equation}

\section{IP Cores on FPGA }
The costs of each operation required to compute the optimal Ate pairing, as presented in this work, are outlined in Table \ref{tab01}. The table includes notations such as $\{a,m,s,i : F_p\}$ and $\{a_2,m_2,s_2,i_2 : F_{p^2}\}$ for operations such as modular addition, subtraction, multiplication, squaring and inversion, as well as $m_\beta$ for multiplication by a constant in $F_p$. Many pairing functions rely on the Miller Loop and Final Exponentiation, which necessitate arithmetic operations in $F_{p^k}$.

\begin{table*} [!t]
\caption{The cost of computing optimal Ate pairing operations}
\label{tab01}
\begin{center}
\color{black}
\begin{tabular}{|c|c|c|c|c|}
\hline
\textbf{Optimal Ate} & \textbf{Add/sub} & \textbf{Multiplication} & \textbf{Squaring} & \textbf{Inversion} \\ \hline
  
\multirow{2}{*}{$F_p$} & \multirow{2}{*}{$a$} & \multirow{2}{*}{$m$} & \multirow{2}{*}{$s$} & \multirow{2}{*}{$i$} \\ 
  &  &  &  &  \\ \hline

\multirow{2}{*}{$F_{p^2}$} & \multirow{2}{*}{$a_2=2a$} & \multirow{2}{*}{$m_2=3m+m_\beta+5a$} & \multirow{2}{*}{$s_2=2m+2m_\beta+5a$} & \multirow{2}{*}{$i_2=4m+m_\beta+2a+i$}  \\ 
  &  &  &  &  \\ \hline

\multirow{2}{*}{$F_{p^6}$} & \multirow{2}{*}{$3a_2$} & \multirow{2}{*}{$6m_2+2m_\beta+15a_2$} & \multirow{2}{*}{$2m_2+3s_2+2m_\beta+10a_2$} & \multirow{2}{*}{$9m_2+3s_2+4m_\beta+5a_2+i_2$} \\ 
  &  &  &  &  \\ \hline

\multirow{2}{*}{$F_{p^{12}}$} & \multirow{2}{*}{$6a_2$} & \multirow{2}{*}{$18m_2+7m_\beta+60a_2$} & \multirow{2}{*}{$12m_2+6m_\beta+45a_2$} & \multirow{2}{*}{$25m_2+9s_2+13m_\beta+61a_2+i_2$} \\ 
  &  &  &  &  \\ \hline

\multirow{2}{*}{$G_{\phi_6}(F_{p^2})$} & \multirow{2}{*}{$6a_2$}    & \multirow{2}{*}{$18m_2+7m_\beta+60a_2$} & \multirow{2}{*}{$6m_2+6m_\beta+39a_2$} & \multirow{2}{*}{Conjugation}  \\ 
  &  &  &  &  \\ \hline

\hline
\hline

\multicolumn{3}{|r|}{\textbf{Sparse multiplication}} & \multicolumn{2}{l|}{$13m_2+3m_\beta+28a_2$  } \\ \hline

\multicolumn{3}{|r|}{\textbf{Doubling and tangent line step}} & \multicolumn{2}{l|}{$3m_2+8s_2+25a_2+4m$  } \\ \hline

\multicolumn{3}{|r|}{\textbf{Addition and line step}} & \multicolumn{2}{l|}{$7m_2+8s_2+25a_2+4m $  } \\ \hline

\end{tabular}
\end{center}
\end{table*}

In this work, we have proposed a technique to perform mathematical operations in the fields $F_{p^6}$ and $F_{p^{12}}$ using arithmetic in the fields $F_p$ and $F_{p^2}$ as outlined in Table \ref{tab01}. This method enables us to avoid the challenge of routing where operations in $F_{p^6}$ and $F_{p^{12}}$ are implemented in hardware. Our approach is intended to minimize resource consumption and to increase system flexibility by working in $F_p$ and $F_{p^2}$. Additionally, we have developed modular operations in both fields, $F_p$ and $F_{p^2}$ as VHDL IP cores, which are controlled by MicoBlaze(s). Furthermore, any curves that require arithmetic in $F_p$ and $F_{p^2}$ can utilize these IP cores by configuring only the software aspect.

\subsection { MMM Core }
The multiplication operation in the base field $F_p$ is a crucial step in computing a cryptographic pairing. There are various methods that can be used to perform this operation. In this paper, we utilize the Montgomery modular multiplication (MMM) algorithm, which is an efficient technique for performing modular multiplication. This algorithm eliminates the need for division by converting modulus reduction into a series of additions and right shifts. The MMM algorithm based on High Radix-$r$ $(r=2^n)$ is defined by the following expression:

$$S_e = Mont(A,B) = (A \times B \times R^{-1}) \; mod \; p$$

$R$ is the Montgomery constant. Algorithm \ref{alg02} illustrates the Montgomery modular multiplication in Radix-$2^{32}$ as presented in \cite{issad2014software}. It is composed of two nested loops $(i)$ and $(j)$. The outer loop $(i)$ is used to calculate the $q_i$ digits. The inner loop $(j)$ incorporates the digits $B[j]$ and $p[j]$ to compute the digits of the intermediate result $S[j-1]$. The final output $S_e$ is obtained when $i=j=e$.

\begin{algorithm}
\caption{Radix-$2^{32}$ Montgomery modular multiplication}
\label{alg02}
\KwData{$A=\sum_{i=0}^e A[i] \times 2^{i\times32}, \;\; B=\sum_{i=0}^e B[i]\times2^{i\times32}, \;\;p=\sum_{i=0}^e p[i]\times2^{i\times32}, $\\
$ \text{Varaibles: }H_i=\sum_{j=0}^e H[j]_i \times 2^{j\times32}, \;\;H1_i=\sum_{j=0}^e H1[j]_i \times 2^{j\times32}, \;\;H2_i=\sum_{j=0}^e H2[j]_i \times 2^{j\times32},$\\
$ \;\;\;\;\;\;\;\;\;\;\;\;\;\;\;C1_i=\sum_{j=0}^e C1[j]_i \times 2^{j\times32}, \;\;C2_i=\sum_{j=0}^e C2[j]_i \times 2^{j\times32},\;\;c1_j=c2_j=c3_j=c4_j,$
$\text{Pre-computed: } p'=-p[0]^{-1} mod 2^{32}$
}
\KwResult{$S_e=\sum_{j=0}^e S[j]_e \times 2^{j\times32}=(A\times B\times R^{-1}) \;mod \; p$}
\SetAlgoLined

$S_0=\sum_{j=0}^e S[j]_0 \times 2^{j\times32}=0$ \\
\For{$i \gets 0$ \textbf{to} $e$} {
    $C1[-1]_i=0; \;\; C2[-1]_i=0$ \\
    $c1_{-1}=c2_{-1}=c3_{-1}=c4_{-1}=0$ \\ 
    $H[0]_i=S[0]_i+A[i]\times B[0]$ \\
    $q_i=(H[0]_i \times p') \; mod \; 2^{32} $ \\
    \For{$j \gets 0$ \textbf{to} $e$} {
        $(C1[j]_i 2^{32},H1[j]_i)=A[i]\times B[j]$ \\
        $(c2_j 2^{32},c1_j 2^{32},H[j]_i)=H1[j]_i+C1[j-1]_i+S[j]_i c1_{j-1}+c2_{j-1}$ \\
        $(C2[j]_i 2^{32},H2[j]_i)=q_i \times p[i]$ \\
        $(c4_j 2^{32},c3_j 2^{32}, S[j-1]_i)=H[j]_i+H2[j]_i+C2[j-1]_i+c3_{j-1}+c4_{j-1}$ \\
    }
    $S[e]_i=c1_e+c2_e+c3_e+c4_e+c1[e]_i+c2[e]_i   $ \\
}
$return \;S_e$ 
\end{algorithm}

For practical use, each operand must be converted to its Montgomery form, adding an extra modular multiplication step due to the $R^{-1}$ factor needed for each multiplication. But in pairing computation, where multiple multiplications occur, the operands only need to be converted once at the start and then back at the end.
                        
\begin{figure*}[t!]
\begin{center}
\includegraphics[width=1\textwidth]{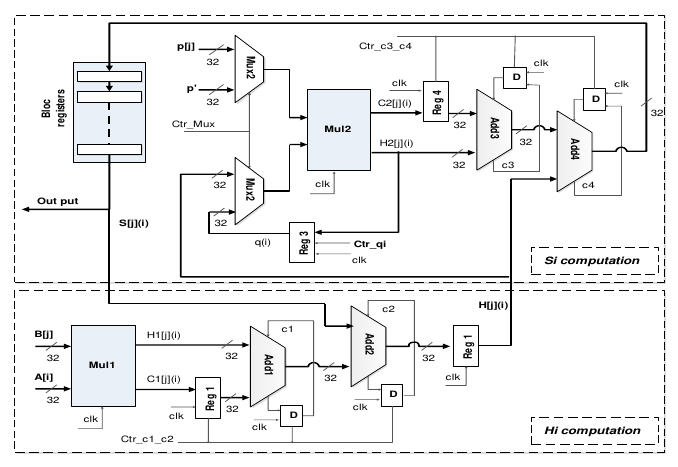} 
\caption{Design of Montgomery Modular Multiplication on an FPGA}
\label{fig01}
\end{center}
\end{figure*}
Our hardware implementation of the Montgomery modular multiplication (MMM) is shown in Figure \ref{fig01}. It follows the operations defined in algorithm \ref{alg02}. The architecture features two 32-by-32 bit multipliers (Mul1 and Mul2), four carry-propagate adders (Add1, Add2, Add3 and Add4), four registers (Reg1, Reg2, Reg3, Reg4), four D Flip-flops, two multiplexers (Mux1, Mux2), and one block register. The inputs $A$, $B$, and $p$ are stored in memory, and the algorithm's intermediate results $S[j]_i$ are stored in the block register as a queue. The MMM core is controlled by four signals: $Ctr\_Mux$, $Ctr\_q_i$, $Ctr\_c1\_c2$, and $Ctr\_c3\_c4$.    

In our implementation of the Montgomery modular multiplication (MMM), we employ the steps in Algorithm \ref{alg02} and the hardware architecture depicted in Figure \ref{fig01}. This architecture encompasses components like multipliers (Mul1 and Mul2), adders, registers, D Flip-flops, multiplexers, and a block register. The execution of the MMM involves storing the operands $A$, $B$, and $p$ in memory, and the intermediate results $S[j]_i$ are temporarily stored in the block register as a queue. The MMM process occurs in three stages: First, the digit $q_i$ is computed and kept in Reg3, which is managed by the signal $Ctr\_q_i$. Then, the multiplications outlined in lines 8, 9 and 10, 11 of Algorithm \ref{alg02} are performed, enabling the computation of the digits $H[j]_i$ and $S[j]_i$. Note that the multiplier Mul2 is shared between the multiplications of lines 6 and 10 in Algorithm \ref{alg02}.

\subsection { KARATSUBA Core }
The arithmetic operations in $F_{p^2}$, including modular addition, subtraction, multiplication, squaring, multiplication by a constant, reduction and inversion, are represented by two numbers in $F_p$. The traditional method of performing modular multiplication in $F_{p^2}$, as outlined in algorithm \ref{alg03}, requires a minimum of four multiplications and five additions/subtractions in $F_p$. However, it can be optimized through parallel computation, which reduces the number of required operations to two multiplications and two additions/subtractions in $F_p$. However, this optimization comes at the cost of duplicating the area required.

\begin{algorithm}
\caption{Karatsuba multiplication method in $F_{p^2}$}
\label{alg03}
\KwData{$A=a_0+a_1\mu, \;\; B=b_0+b_1\mu$}
\KwResult{$C=c_0+c_1\mu$}
\SetAlgoLined
\DontPrintSemicolon

$t_0 \gets a_0*b_0, \;\;   tmp \gets b_0+b_1 $ \;
$t_1 \gets a_1*b_1, \;\;   c_1 \gets a_0+a_1 $ \;
$c_0 \gets t_1*redFp$ \;
$c_1 \gets c_1*tmp$ \;
$c_0 \gets c_0-t_0$ 
\end{algorithm}

\begin{figure*}[!h]
\begin{center}
\includegraphics[scale=0.75]{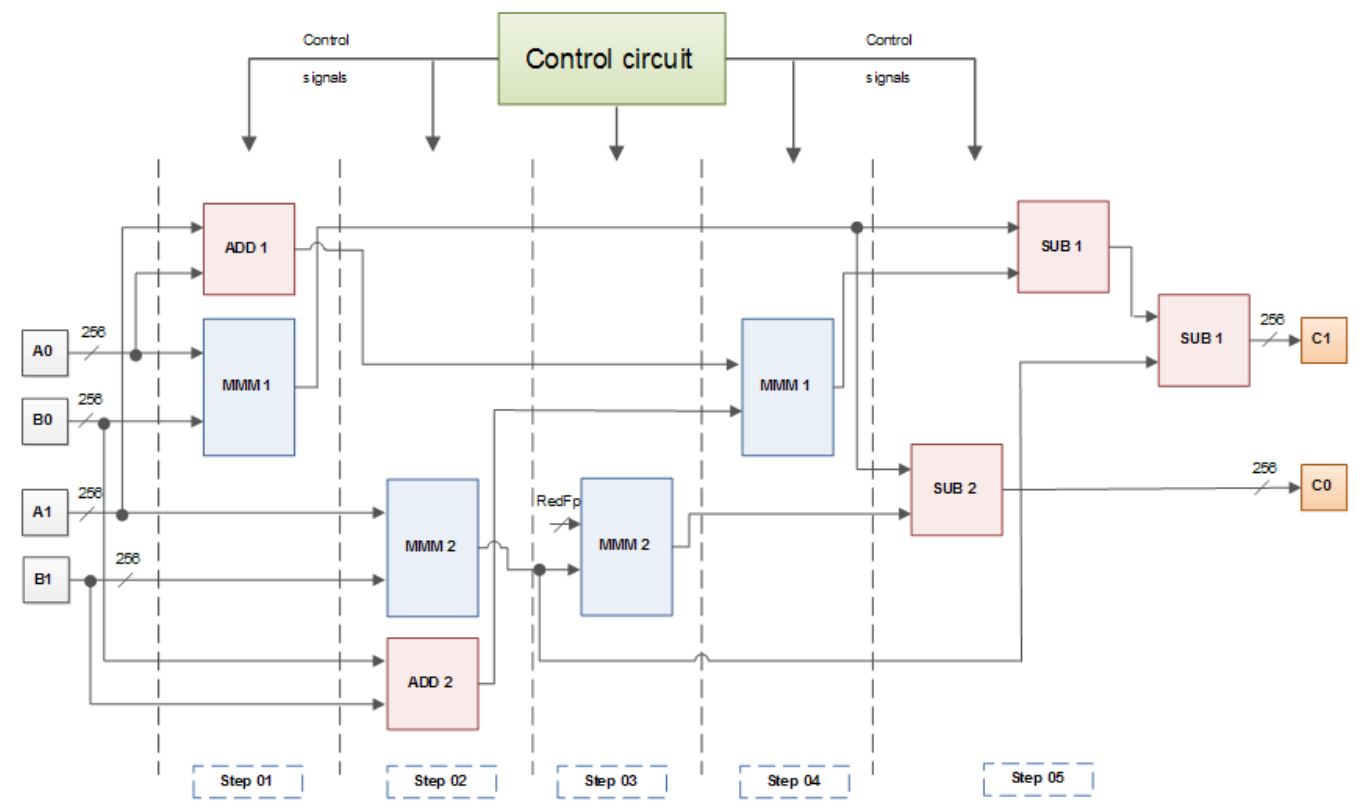} 
\caption{Hardware architecture of KARATSUBA on FPGA}
\label{fig02}
\end{center}
\end{figure*}

In this work, the KARATSUBA IP core is introduced, which facilitates the performance of various modular operations in $F_{p^2}$, including multiplication, squaring, constant multiplication, and reduction. Furthermore, it can be employed to execute Montgomery modular multiplication in $F_p$. The design of KARATSUBA is shown in Figure \ref{fig02} and involves five stages, controlled by a control circuit that selects the appropriate IPs for each stage. The core includes the MMM and ADD/SUB IP VHDL cores. This later is used for modular addition/subtraction in $F_p$.

\textcolor{black}{
The KARATSUBA IP enhances the efficiency of modular operations in $F_{p^2}$. All of these operations can be executed by a single IP, resulting in reduced computational resources. The KARATSUBA IP leverages the Karatsuba algorithm, a fast multiplication technique, to minimize the number of elementary multiplications compared to traditional methods. Moreover, it improves the execution time compared to purely software implementations. In particular, the computation cost of modular multiplication in $F_{p^2}$ was initially 53942 cycles with pure Software implementation in MicroBlaze, but it reduced significantly to only 1240 cycles with the use of KARATSUBA IP. Table \ref{tab02} presents the FPGA-based hardware outcomes of the KARATSUBA IP.
}

\begin{table} [!h]
\caption{\textcolor{black}{The hardware results of KARATSUBA}}
\label{tab02}
\begin{center}
\color{black}
\begin{tabular}{|c||c|c|c||c|}
\hline

\textbf{IP Core} & \textbf{Slices} & \textbf{DSP} & \textbf{BRAM} & \textbf{cycles} \\ \hline

\textbf{ADD/SUB} & 487 & 6 & 7 & 10 \\ \hline

\textbf{MMM} & 495 & 8 & 3 & 130 \\ \hline

\textbf{KARATSUBA} & 982 & 14 & 10 & 550 \\ \hline

\end{tabular}
\end{center}
\end{table}

\section{ Proposed architectures for optimal Ate pairing }
In this research, we propose three different designs for implementing the optimal Ate pairing algorithm as an embedded system on an FPGA. We will now describe these hardware architectures in detail.

\subsection{Signal MicroBlaze-based software implementation}
In this approach, a fully software-based implementation of optimal Ate on a Genesys board is presented as a pioneering solution. The method involves storing all the required functions and operations for computing Optimal Ate in memory (BRAM) and executing them sequentially using a MicroBlaze processor.

The MicroBlaze is a 32-bit RISC soft processor designed by Xilinx for embedded systems, and can be implemented on various development boards from Xilinx or their partners. It offers fundamental operations like addition, subtraction, and multiplication. To achieve optimal performance in Ate pairing, all operands are represented in 32-bit packets.

\begin{figure*}[!h]
\begin{center}
\includegraphics[width=0.75\textwidth]{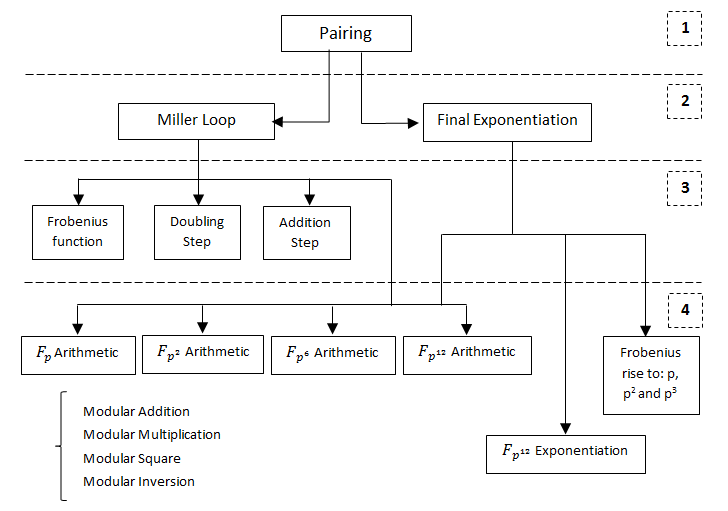} 
\caption{Optimal Ate pairing implementation hierarchy}
\label{fig03}
\end{center}
\end{figure*}

The optimal Ate pairing is executed through a C program on the MicroBlaze using SDK tools. The software architecture is structured into four levels as depicted in Figure \ref{fig03}. The top level encompasses the pairing function, followed by the second level that focuses on the Miller algorithm and Final Exponentiation. The third level consists of the Doubling step, Addition step, and the Frobenius function. Finally, the fourth level encompasses the Frobenius operations, arithmetic operations in finite fields (such as addition, subtraction, multiplication, and division), and exponentiation in fields $F_p$, $F_{p^2}$, $F_{p^6}$, and $F_{p^{12}}$.

Most operations are performed in the extended fields of quadratic $(F_{p^2})$ and cubic $(F_{p^3})$ within the towering scheme $F_{(((p^2)^2)^3)}$. According to \cite{devegili2006multiplication}, there are several techniques available for multiplication and squaring in such extended fields. In particular, the Karatsuba approach is utilized for multiplication and the complex method is implemented for squaring in $F_{p^2}$. Additionally, the Karatsuba method is applied for both multiplication and squaring in $F_{p^3}$.

\begin{figure}[!h]
\begin{center}
\includegraphics[width=8.4cm, height=6cm]{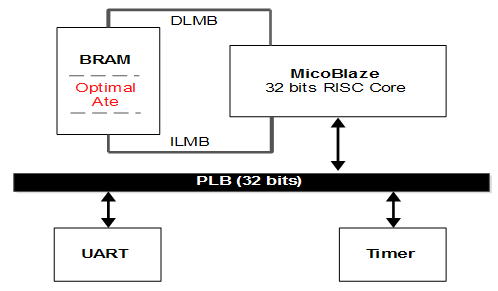} 
\caption{Hardware architecture of Mb software approach}
\label{fig04}
\end{center}
\end{figure}

The hardware architecture for executing the optimal Ate pairing on a Virtex-5 circuit with a MicroBlaze processor is illustrated in Figure \ref{fig04}. The design encompasses a MicroBlaze processor, Block Random Access Memory (BRAM), Local Memory Buses (ILMB, DLMB) to organize the BRAM, a Timer for timing the execution, and a Universal Asynchronous Receiver Transmitter (UART) to communicate input and output data with the serial port.

\subsection{Single MicroBlaze-based SW/HW implementation}
The second approach in this work involves a combination of software and hardware design for optimal Ate pairing on BN-curves using a Virtex-5 circuit. To enhance performance, an accelerator IP core was integrated into the design and implemented in conjunction with the MicroBlaze processor. \textcolor{black}{This approach aims to improve the overall execution time compared to the initial one. In particular, the computation cost of modular multiplication in $F_p$ was initially 12968 cycles with pure software implementation in MicroBlaze, but it reduced significantly to only 475 cycles with the use of MMM IP.}

\begin{figure}[!h]
\begin{center}
\includegraphics[width=8.4cm, height=6.cm]{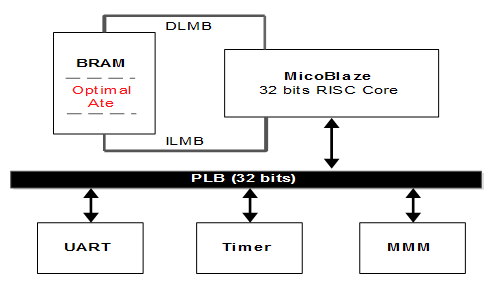} 
\caption{Hardware architecture of Mb/MMM approach}
\label{fig05}
\end{center}
\end{figure}

The first design based on this approach use our MMM core to perform all the necessary modular multiplication operations, which can be significant for pairing defined on 256-bit BN-curves, as shown in reference \cite{aranha2011faster}. The hardware architecture for this approach is illustrated in Figure \ref{fig05}.

\begin{figure}[!h]
\begin{center}
\includegraphics[width=8cm, height=6cm]{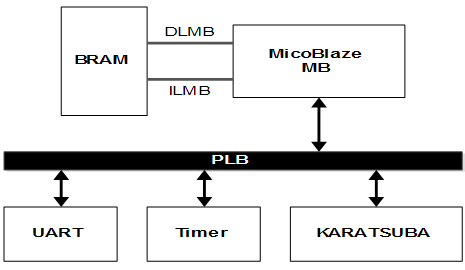} 
\caption{Hardware architecture of Mb/KARATSUBA approach}
\label{fig06}
\end{center}
\end{figure}

In the second design of the hardware/software approach for optimal Ate pairing on BN-curves, a KARATSUBA core is utilized in conjunction with the MicroBlaze processor to perform all necessary operations in the fields $F_p$ and $F_{p^2}$. The architecture of this embedded system is depicted in Figure \ref{fig06}.

The partitioning method suggested in this work combines both software and hardware elements, resulting in improved execution speed and increased flexibility in the design of the embedded system. The higher level functions are implemented in software, while the lower level functions are executed by specialized IP cores. However, it is important to note that the transfer time of data between the MicroBlaze and the IP core also plays a significant role in determining the overall execution time.

\begin{figure}[!h]
\begin{center}
\includegraphics[width=0.75\columnwidth]{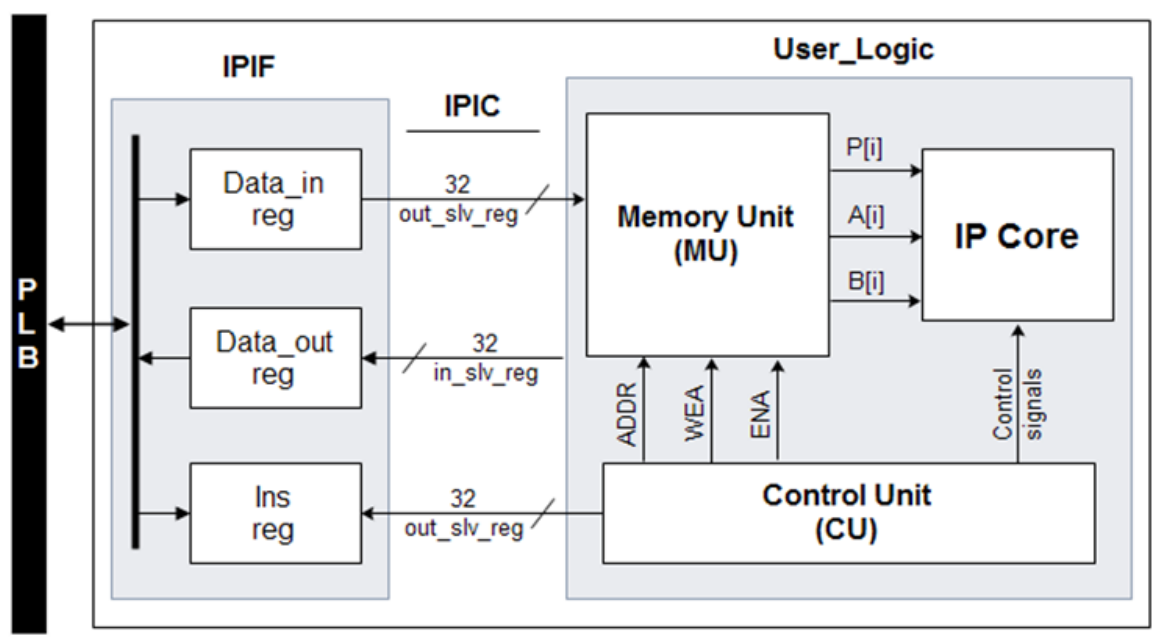} 
\caption{The design of the hardware components for our IP cores.}
\label{fig07}
\end{center}
\end{figure}

The overall structure of how our IP cores are integrated with the MicroBlaze processor is depicted in Figure \ref{fig07}. The IP cores are connected to the MicroBlaze through the use of the Xilinx PLB Bus, which facilitates the exchange of data and instructions between the two components. The design includes the Xilinx Intellectual Property InterFace (IPIF) and User Logic blocks, which communicate with each other through a standard interface called IP InterConnect (IPIC).

The IP cores' integration with the MicroBlaze processor is shown in Figure \ref{fig07}. The cores are linked to the MicroBlaze through the PLB Bus from Xilinx, which manages data and instruction transfer. The architecture consists of Xilinx's Intellectual Property Interface (IPIF) and User\_Logic blocks, communicating via the standard IP InterConnect (IPIC) back-end interface. The IPIF interface decodes the PLB system bus communication protocol and has three registers: Ins\_reg, DataIn\_reg, and DataOut\_reg. MicroBlaze sends instruction codes via the Ins\_reg instruction register. The User\_Logic block implements the circuit logic and includes three units: the Memory Unit (MU), the Control Unit (CU), and the IP core. The CU retrieves instructions from the Ins\_reg and manages the MU and IP core.

\subsection{Dual MicroBlaze-based SW/HW implementation}
\color{black} 
Optimal Ate pairing demonstrates parallelism at various levels, ranging from functions in $F_p$ to higher levels in $F_{p_{12}}$. As we move from lower to higher levels, a significant level of parallelism becomes evident, providing the impetus for exploring and developing diverse architecture configurations. These configurations involve variations in hardware components, including the number of MicroBlaze processors and KARATSUBA IPs employed.

In our study, we have investigated multiple software/hardware architecture configurations for the implementation of optimal Ate pairing. This analysis enables us to evaluate the performance and hardware resources utilized in each configuration, aiding in the identification of the most resource-efficient option while maintaining reasonable execution time. Several architectures can be explored and developed, such as: 1MB/2KARATSUBA, 1MB/3KARATSUBA, 2MB/1KARATSUBA, 2MB/2KARATSUBA, 3MB/1KARATSUBA, 3MB/2KARATSUBA, 3MB/3KARATSUBA, and more.

The third approach, in this work, focuses on utilizing the inherent parallelism of key operations, which include modular multiplication in $F_{p^6}$ and $F_{p^{12}}$, sparse multiplication, squaring in the cyclotomic subgroup $G_{\phi6}(F_{p^2})$, as well as doubling and addition steps. Moreover, parallelism becomes crucial when executing frequently repeated key operations for calculating optimal Ate. For example, algorithm \ref{alg04} shows the multiplication function in $F_{p^6}$.

\begin{algorithm}[h]
\caption{\textcolor{black}{Multiplication in $F_{p^6}$}}
\color{black}
\label{alg04}
\KwData{$A=a_0+a_1 x+a_2 x^2, \;\; B=b_0+b_1 x+b_2 x^2$}
\KwResult{$C=c_0+c_1 x+c_2 x^2$}
\SetAlgoLined
\DontPrintSemicolon

$t_0 \gets a_0 * b_0$ \;
$t_1 \gets a_1 * b_1$ \;
$t_2 \gets a_2 * b_2$ \;
$c_0 \gets [((a_1+a_2 )*(b_1+b_2 ))-t_1-t_2 ].\xi +t_0$ \;
$c_1 \gets [((a_0+a_1 )*(b_0+b_1 ))-t_0-t_1 ]+t_2 .\xi$ \;
$c_2 \gets [((a_0+a_2 )*(b_0+b_2 ))-t_0-t_2 ]+t_1$\;

\end{algorithm}

The cost of algorithm \ref{alg04} is : 6 Karatsuba + 15 add $F_{p^2}$ + 2 red $F_{p^2}$

After developing and testing the various operations/functions on the Virtex5 board, We have obtained the following significant result.

$$
\textbf{2 add soft $F_{p^2}$ (1272 cycles)} >\approx \textbf{Karatsuba $F_{p^2}$ (1240 cycles)} 
$$
$$
\textbf{add soft $F_{p^2}$ (636 cycles)} \approx \textbf{red $F_{p^2}$ (590 cycles)}
$$

We have the execution time of a single multiplication in $F_{p^2}$, which is almost the same as that of two addition operations in software on MicroBlaze. Additionally, the execution time of a reduction operation in $F_{p^2}$ using KARATSUBA is almost the same as that of a software addition operation on MicroBlaze.
Based on these results, the algorithm \ref{alg04} is executed on the two processors, as it shown in table \ref{tab03}.

\begin{table*} [!t]
\caption{\textcolor{black}{Multiplication in $F_{p^6}$ (2Mb/KARATSUBA)}}
\label{tab03}
\begin{center}
\color{black}
\begin{tabular}{|c|c|c||c|}
\hline

\textbf{MB0 (master)~~~~~~~~~~~~~~~~~~} & \textbf{~~~~~~~~~~~Transfer FSL~~~~~~~~~~~} & \textbf{~~~~~~~~~~~MB1 (slave)~~~~~~~~~~~} & \textbf{~~~~~~~~~~~Cost~~~~~~~~~~~~~}  \\ \hline

- & $\{a_0,b_0\}$ & - & 2t  \\ \hline

$ta_{01} \gets a_0+a_1$ & \multirow{2}{*}{-}  &  \multirow{2}{*}{$t_0 \gets a_0*b_0$} & \multirow{2}{*}{2 add $F_{p^2}$}  \\ 
$tb_{01} \gets b_0+b_1$ &   &   &   \\ \hline

- & $\{a_1,b_1\}$ & - & 2t  \\ \hline

$ta_{02} \gets a_0+a_2$ & \multirow{2}{*}{-}  &  \multirow{2}{*}{$t_1 \gets a_1*b_1$} & \multirow{2}{*}{2 add $F_{p^2}$}  \\ 
$tb_{01} \gets b_0+b_1$ &   &   &   \\ \hline

- & $\{a_2,b_2\}$ & - & 2t  \\ \hline

$ta_{12} \gets a_1+a_2$ & \multirow{2}{*}{-}  &  \multirow{2}{*}{$t_2 \gets a_2*b_2$} & \multirow{2}{*}{2 add $F_{p^2}$}  \\ 
$tb_{12} \gets b_1+b_2$ &   &   &   \\ \hline

- & $\{ta_{12},tb_{12}\}$ & - & 2t  \\ \hline

- & - & $ta_{12} \gets ta_{12}*tb_{12}$ & karatsuba  \\ \hline

- & $\{ta_{01},tb_{01},ta_{12},t_1,t_2\}$ & - & 2t+3r  \\ \hline

$ta_{12} \gets ta_{12}-t_1$ & \multirow{2}{*}{-}  &  \multirow{2}{*}{$ta_{01} \gets ta_{01}*tb_{01}$} & \multirow{2}{*}{2 add $F_{p^2}$}  \\ 
$ta_{12} \gets ta_{12}-t_2$ &   &   &   \\ \hline

- & $\{ta_{02},tb_{02},ta_{01},t_0\}$ & - & 2t+2r  \\ \hline

$ta_{01} \gets ta_{01}-t_0$ & \multirow{2}{*}{-}  &  \multirow{2}{*}{$ta_{02} \gets ta_{02}*tb_{02}$} & \multirow{2}{*}{2 add $F_{p^2}$}  \\ 
$ta_{01} \gets ta_{01}-t_1$ &   &   &   \\ \hline

- & $\{ta_{12},ta_{02}\}$ & - & 1t+1r  \\ \hline

$ta_{02} \gets ta_{02}-t_0$ & \multirow{2}{*}{-}  &  $ta_{12} \gets ta_{12}.\xi$ & \multirow{2}{*}{2 add $F_{p^2}$}  \\ 
$ta_{02} \gets ta_{02}-t_2$ &   & $tb_{01} \gets t_2.\xi$  &   \\ \hline

- & $\{tb_{01}\}$ & - & 1r  \\ \hline

$c_1 \gets ta_{01}+tb_{01}$ & \multirow{2}{*}{-}  & \multirow{2}{*}{$c_0 \gets ta_{12}+t_0$}  & \multirow{2}{*}{2 add $F_{p^2}$}  \\ 
$c_2 \gets ta_{02}+t_1$ &   &   &   \\ \hline

- & $\{c_0\}$ & - & 1r  \\ \hline\hline

$90,01\%$ & - & $76,82\%$ & Percentage  \\ \hline

\end{tabular}
\end{center}
\end{table*}

\{$.\xi$\} respresnts modular reduction in $F_{p^2}$. \{t and r\} represent the transfer time by FSL.

Now, the cost of algorithm \ref{alg04} is : Karatsuba + 14 add soft $F_{p^2}$ + 21 transfert FSL.

We can clearly observe a significant improvement in execution time for the multiplication function in $F_{p^6}$. 

The same principle is applied to the key operations/functions in optimal Ate, such as functions in $F_{p^6}$, $F_{p^{12}}$, Doubling and Addition steps, Sparse multiplication, exponentiation, and so on.

\begin{figure}[!h]
\begin{center}
\includegraphics[width=0.9\columnwidth]{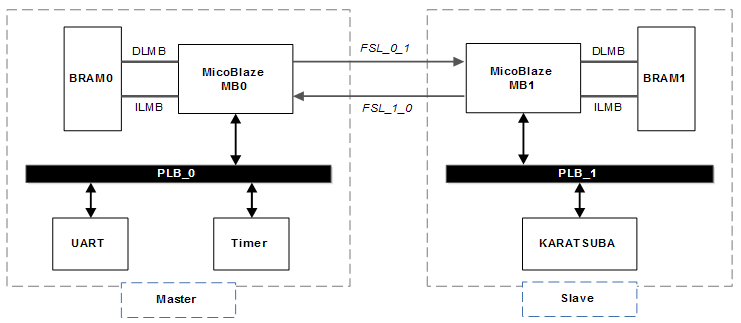} 
\caption{Hardware architecture of 2Mb/KARATSUBA approach}
\label{fig08}
\end{center}
\end{figure}

In order to implement optimal Ate pairing in an efficient manner, a specific architecture was chosen that meets the criteria of minimal memory usage while maintaining an acceptable execution time. As shown in Figure \ref{fig08}, a parallel and flexible approach is proposed, utilizing two MicroBlaze processors and an IP KARATSUBA core. The processors, labeled as $MB_0$ and $MB_1$, are connected through a high-speed FSL bus. The $MB_0$ processor acts as the master and $MB_1$ acts as the slave, responsible for performing operations in $F_p$ and $F_{p^2}$ in conjunction with the KARATSUBA core, which is connected through a PLB bus.

The first idea involves adding KARATSUBA IPs around a single MicroBlaze processor. However, this approach has a major drawback, which is the transfer time between the MicroBlaze processor and the different IPs. To illustrate, conducting a multiplication operation in $F_{p^2}$ demands a total of 1240 cycles, with 550 cycles allocated for processing and an additional 690 cycles dedicated to data transfer. Notably, the processing time nearly matches the transfer time, underscoring the crucial role of transfer time in the overall system performance. Additionally, the MicroBlaze and the IP cannot work at the same time. The MicroBlaze always waits for the results sent by the IP. Consequently, architectures with parallel processors have emerged as a more favored alternative.

To determine the percentage of task between the MicroBlaze and the IP core, we can achieve this by thoroughly examining the algorithms involved in key operations, such as multiplication in $F_{p^6}$ and $F_{p^{12}}$, squaring in $G_{\phi6}F_{p^2}$, sparse multiplication, doubling step, and others. Through this analysis, we can determine the specific tasks or operations allocated to each component and evaluate their respective contributions.For instance, the multiplication in $F_{p^6}$ is executed with a percentage of 90.01\% on the first processor and 76.82\% on the second, as shown in Table \ref{tab03}.

\color{black}

\section{Implementation Results and Discussion}
\subsection{\textcolor{black}{Implementation Results}}
The design of an embedded system for computing optimal Ate pairings was created using the Xilinx Platform Studio environment and a Virtex-5 Genesys development board. The MMM and KARATSUBA components were written in VHDL and tested with Modelsim SE before being synthesized with the ISE Design Suite. The DSP48E and RAM blocks were created with the Core Generator tool, while the high-level arithmetic was developed with C programming language in the SDK.

In order to ensure that the proposed design offers a 128-bit security level, we selected the parameter $t = 2^{62}-2^{54}+2^{44}$, as stated in \cite{beuchat2010high}, and the BN-curve $E: y^2=x^3+5$. This choice results in the exponent number $t$ and the parameter $s=6t+2$ in the Miller Loop having a signed bit length of 63 and 65, respectively.

A comparison of the results obtained from our implementation of optimal Ate pairing with those of recent implementations based on BN-curves is presented in Table \ref{tab04}. The comparison takes into account execution time, hardware requirements, and design efficiency, which is computed using a this expression:

\begin{equation}
\mathtt{efficiency}=\frac{\mathtt{datapath(bit)}}{\mathtt{occupiedarea(slice)%
}\times \mathtt{executiontime(s)}}
\end{equation}

The area of the design is determined by taking into consideration the following information: it is assumed that the height of a DSP48E is equivalent to that of five configurable logic blocks (CLBs) and is also equivalent to the height of one block RAM. Each CLB is made up of four slices.

\textcolor{black}{The number of BRAMs is configurable on Xilinx FPGA boards. By default, when working on a new project, the number of BRAMs is set to 18. However, our initial implementation (purely software-based) required us to increase this number to 32. However, with the task separation and the utilization of two MicroBlaze processors, this number increased to 42.}

The implementation of optimal Ate pairing through software running on MicroBlaze has a slower speed in comparison to other approaches. While the SW/HW design aims to optimize the execution time, it also results in an increased consumption of hardware area. The transfer time between the MicroBlaze processor and IP cores can also affect the global execution time. Taking advantage of the parallel elements in optimal Ate pairing holds potential for improving the global execution time while keeping hardware usage minimal.
\textcolor{black}{Table \ref{tab04} summarizes the evaluation of the cost associated with key functions in Optimal Ate utilizing 2Mb/KARATSUBA.}

\begin{table*} [!t]
\caption{Evaluating the cost of key functions in Optimal Ate with 2Mb/KARATSUBA}
\label{tab04}

\begin{center}
\color{black}
\begin{tabular}{|c|c|c|c|c|}
\hline

\multirow{2}{*}{\textbf{Functions}} & \multirow{2}{*}{\textbf{Design}} & \multirow{2}{*}{\textbf{The cost}} & \multicolumn{2}{c|}{\textbf{Percentage of task}} \\ \cline{4-5}
  &  &  & $MB_0$ & $MB_1$ \\ \hline

\multirow{2}{*}{Mult $F_{p^6}$} & Mb/KARATSUBA & 6 Karatsuba + 15 add soft $F_{p^2}$ + 2 red $F_{p^2}$ & \multicolumn{2}{c|}{100\%} \\ \cline{2-5}
& 2Mb/KARATSUBA & Karatsuba + 14 add soft $F_{p^2}$ + 21 transfert FSL & 90.01\% & 76.82\%\\ \hline

\multirow{2}{*}{Mult $F_{p^{12}}$} & Mb/KARATSUBA & 18 Karatsuba + 60 add soft $F_{p^2}$ + 7 red $F_{p^2}$ & \multicolumn{2}{c|}{100\%} \\ \cline{2-5}
& 2Mb/KARATSUBA & Karatsuba + 51 add soft $F_{p^2}$ + 68 transfert FSL & 97,04\% & 75.46\%\\ \hline

\multirow{2}{*}{Squaring in $G_{\phi6}F_{p^2}$} & Mb/KARATSUBA & 6 Karatsuba + 39 add soft $F_{p^2}$ + 6 red $F_{p^2}$  & \multicolumn{2}{c|}{100\%} \\ \cline{2-5}
& 2Mb/KARATSUBA & Karatsuba + 27 add soft $F_{p^2}$ + 27 transfert FSL & 93.16\% & 82.98\%\\ \hline

\multirow{2}{*}{Sparce multiplication} & Mb/KARATSUBA & 14 Karatsuba + 28 add soft $F_{p^2}$ + 3 red $F_{p^2}$ & \multicolumn{2}{c|}{100\%} \\ \cline{2-5}
& 2Mb/KARATSUBA & 8 Karatsuba + 20 add soft $F_{p^2}$ + 34 transfert FSL & 84.23\% & 78.78\%\\ \hline

\multirow{2}{*}{Doubling step} & Mb/KARATSUBA & 13 Karatsuba + 24 add soft $F_{p^2}$ & \multicolumn{2}{c|}{100\%} \\ \cline{2-5}
& 2Mb/KARATSUBA & Karatsuba + 24 add soft $F_{p^2}$ + 25 transfert FSL & 93.92\% & 79.01\%\\ \hline

\end{tabular}
\end{center}
\end{table*}

\begin{table*} [!t]
\caption{Comparison of results from implementation of Optimal Ate pairing}
\label{tab05}
\begin{center}
\begin{tabular}{|c|c|c|c|c|c|c|c|c|c|}
\hline
\multirow{2}{*}{\textbf{Ref.}} & \multirow{2}{*}{\textbf{Platform}} & \textbf{Freq.} & \multirow{2}{*}{\textbf{Design}} & \multicolumn{3}{c|}{\textbf{Area}} & \multirow{2}{*}{\textbf{Cycles}} & \textbf{Times} & \multirow{2}{*}{\textbf{efficiency}} \\ \cline{5-7}
    &  & MHz  &  & \textbf{Slices} & \textbf{DSP}  & \textbf{RAM}  &  & (ms)   &  \\ \hline

\multirow{8}{*}{Our} & \multirow{8}{*}{Virtex-5} & \multirow{2}{*}{125} & SW  & \multirow{2}{*}{1063} & \multirow{2}{*}{3} & \multirow{2}{*}{32} & \multirow{2}{*}{262445486} & \multirow{2}{*}{2099.56} & \multirow{2}{*}{0.07}    \\ 
 &  &  &  Mb  &  &  &  &  &  &  \\ \cline{3-10}

 &  & \multirow{2}{*}{100}  & SW/HW  & \multirow{2}{*}{1558} & \multirow{2}{*}{11} & \multirow{2}{*}{35}        & \multirow{2}{*}{26551593} & \multirow{2}{*}{265.51}  & \multirow{2}{*}{0.42} \\ 
 &  &   & Mb/MMM &   &   &  &  &   &    \\ \cline{3-10}

 &  & \multirow{2}{*}{100}  & SW/HW  & \multirow{2}{*}{2045} & \multirow{2}{*}{17} & \multirow{2}{*}{38} & \multirow{2}{*}{1122817} & \multirow{2}{*}{112.28} & \multirow{2}{*}{0.77}   \\ 
 &  &  & Mb/KARATSUBA  &    &    &   &    &   &  \\ \cline{3-10}

 &  & \multirow{2}{*}{100}  & SW/HW  & \multirow{2}{*}{3108} & \multirow{2}{*}{20} & \multirow{2}{*}{42} & \multirow{2}{*}{264668} & \multirow{2}{*}{26.4} & \multirow{2}{*}{2.35}\\
 &  &  & 2Mb/KARATSUBA &    &  &  &  &  & \\ \hline

\hline
\cite{ghosh2010high}   & Virtex-4 & 50  & HW & 52000  & -  & - & 821000 & 16.42 & 0.29 \\ \hline
\cite{ghosh2012secure} & Viretx-6 & 145 & HW & 23000  & -  & - & 821000 & 5.66  & 1.96  \\ \hline
\cite{hao2016dual}     & Virtex-5 & 125 & HW & 10592  & 51 & - & 283111 & 2.26  &9.92  \\ \hline
\cite{xie2022high} & Viretx-6 & 72 & HW & 25000  & 240 & - & 37271  & 0.52  & 17.07 \\ \hline
\cite{sghaier2018high} & Viretx-6 & 225 & HW & 5570   & 30 & - & 80000  & 0.35  & 120 \\ \hline

\end{tabular}
\end{center}
\end{table*}

\subsection{\textcolor{black}{Discussion}}
\color{black}
The research presented in the paper contributes to advancing the state-of-the-art in optimal Ate pairing algorithms by offering area-efficient and flexible architectures. These architectures enhance the efficiency, performance, and practicality of cryptographic pairings, opening up possibilities for secure and efficient implementations in various applications such as identity-based cryptography, attribute-based encryption, and cryptographic protocols involving
pairings. The research's key implications and contributions encompass three aspects: (i) The proposal of novel architectures targeting area efficiency in FPGA implementations of optimal Ate pairing, which holds particular relevance for resource-constrained environments necessitating effective FPGA resource utilization. (ii) Emphasizing flexibility, the architectures can be easily tailored and adapted to suit diverse parameters and security requirements of the optimal Ate pairing implementation. (iii) Providing practical insights and experimental findings by conducting FPGA-based implementations and tests, delivering valuable guidance for real-world applications.\\
In the context of this study, the outcomes achieved through the implementation of optimal Ate pairing on FPGA are discussed, and the results are presented in Table \ref{tab05}.

Typically, in \cite{ghosh2010high}, the first implementation of pairing functions for 128-bit security level using BN-curves was reported. The authors utilized Blakley's algorithm for modular multiplication, leading to a high area consumption without using DSP or RAM cores. Our SW/HW designs, on the other hand, show improved slice consumption and efficiency. In \cite{ghosh2012secure}, a fully hardware-based implementation of Ate and optimal Ate pairing was presented, where all $F_{p^k}$-arithmetic was implemented in hardware. This design utilized 23k logic slices, but had a faster time performance compared to the 2Mb/KARATSUBA design. However, it also consumed more slices, being 5.6 times higher.
Moving on, the authors in \cite{hao2016dual} proposed a hardware cryptoprocessor for optimal Ate pairing which utilizes two processing engines to perform parallel computation of $F_p$-arithmetic using the Montgomery algorithm. This design has a reasonable increase in area with a higher number of DSP blocks, making it limited to integration on high-resource FPGA boards, unlike our designs which can be implemented on large FPGA circuits. In \cite{xie2022high}, a high-performance processor for optimal Ate pairing on BN-curves is proposed, exploiting parallelism and pipeline at various levels of the algorithm. However, this design has a higher area occupation with a higher number of DSP blocks and is not suitable for restricted environments.
In \cite{sghaier2018high}, a high-speed and efficient design for optimal Ate pairing over BN and BLS12 curves on FPGA was presented. The design boasts the highest reported speed and the best reported area-time performance. Although the design offers improved efficiency compared to our implementations, it has a reasonable increase in area and is less flexible.

All in all, these findings open avenues for further research and optimization in implementing efficient and secure cryptographic systems.

\color{black}

\section{Conclusion}
In this paper, we proposed three different approaches for implementing optimal Ate pairing based on Jacobean coordinates over BN-curves with 128-bit security as an embedded system on FPGA devices. Our first approach utilized a pure software design executed by MicroBlaze processors, while the second approach combined software and hardware to perform essential operations in $F_p$ and $F_{p^2}$. Our third approach employed parallelism at critical operation levels to further improve execution time and minimize area consumption. Our designs are suitable for restricted environments and offer reasonable execution times.

\color{black}


To further improve the implementation of optimal Ate pairing and address potential limitations, the following aspects can be considered in future works: (i) investigating and implementing algorithmic optimizations to enhance the efficiency of the Ate pairing computation. Research on new techniques or adaptations specific to the BN-curve can lead to significant improvements in execution time and resource utilization; (ii) exploring the use of more advanced FPGA platforms or application-specific integrated circuits (ASICs) to increase computational capabilities and achieve higher performance. Utilizing modern FPGA families with improved hardware resources and higher clock frequencies can result in faster computations; (iii) investigating and incorporating parallelization and pipelining techniques to exploit parallel hardware resources effectively. By distributing tasks across multiple processing units and overlapping computations, the overall execution time can be reduced; (iv) designing and implementing custom hardware accelerators tailored to the specific requirements of the optimal Ate pairing computation. This can lead to dedicated hardware units optimized for the BN-curve operations, further improving efficiency; (v) focusing on optimizing power consumption and resource utilization without compromising security. This is especially important for embedded systems and IoT devices where energy efficiency is a critical consideration;  (vi) exploring opportunities for further software-level optimization, such as using advanced compiler techniques or employing custom assembly code to fine-tune critical arithmetic operations; (vii) Conducting a thorough security analysis of the proposed design against potential side-channel attacks and fault injections. Implement countermeasures to mitigate these vulnerabilities and ensure robustness against various security threats; and (viii) evaluating the implementation in real-world applications, such as secure communication protocols or cryptographic schemes, to assess its practical viability and gather feedback for further improvements.

\color{black}

\section*{Ethical Approval} 
Not Applicable
 
\section*{Competing interests}
The authors declare no conflict of interest.
 
\section*{Authors' contributions}
Conceptualization, O. Azzouzi - Methodology, O. Azzouzi, M. Anane - Validation,  O. Azzouzi, M. Anane, M. Koudil - Writing—original draft preparation, O. Azzouzi, M. Issad, Y. Himeur - Proofreading, M. Anane, Y. Himeur - Formal analysis, M. Anane, M. Koudil, M. Issad, Y. Himeur - Supervision, M. Anane, M. Koudil- Project administration, M. Anane.

\section*{Funding}
This research received no external funding

\section*{Availability of data and materials}
Data will be shared upon request



\end{document}